# Thermally Activated Asymmetric Structural Recovery in a Soft Glassy Nano-Clay Suspension

Tanmay P. Dhavale, Shweta Jatav and Yogesh M Joshi*
Department of Chemical Engineering, Indian Institute of Technology Kanpur, Kanpur 208016. INDIA.
* E-Mail: joshi@iitk.ac.in

**Abstract**

In this work we study structural recovery of a soft glassy Laponite suspension by monitoring temporal evolution of elastic modulus under isothermal conditions as well as following step temperature jumps. Interestingly, evolution behavior under isothermal conditions indicates the rate, and not the path of structural recovery, to be dependent on temperature. The experiments carried out under temperature jump conditions however trace a different path of structural recovery, which shows strong dependence on temperature and the direction of change. Further investigation of the system suggests that this behavior can be attributed to restricted mobility of counterions associated with Laponite particle at the time of temperature change, which do not allow counterion concentration to reach equilibrium value associated with the changed temperature. Interestingly this effect is observed to be comparable with other glassy molecular and soft materials, which while evolve in a self-similar fashion under isothermal conditions, show asymmetric behavior upon temperature change.

## I. Introduction

Glassy materials, such as molecular and soft (colloidal) glasses, are out of thermodynamic equilibrium owing to constrained mobility of the trapped constituting entities. In molecular glasses kinetically arrested state is achieved by rapidly decreasing the temperature, while in soft glasses the same is obtained by rapidly increasing the concentration of constituents or by altering the inter-particle interactions.[1] Under quiescent conditions, these materials undergo structural organization as a function of time. This process is known as physical aging or structural recovery.[2-4] During this process, the molecular glasses typically undergo specific volume relaxation (or densification) as a function of time. Aging or structural recovery in soft glasses, on the other hand, can be represented merely as progressive reduction in free energy as a function of time, as there is no generic structural variable associated with this process.[5] In any glassy material, temperature is an important variable that affects the timescale of structural organization, and its effect on molecular glasses has been studied in great details.[6] In this work we study how physical aging in soft glassy nano-clay (Laponite RD) suspension is affected by change in temperature, and the similarity it shares with structural recovery in molecular and other colloidal glasses.

In order to have physical aging in the materials arrested in thermodynamically out of equilibrium state, the trapped entities must possess sufficient kinetic (thermal) energy to carry out structural organization. The extreme case of total absence of thermal energy in the trapped entities is randomly packed stationary granular media, wherein the aging is completely absent, and the system remains arrested in the high energy state forever. Enhancement in thermal energy caused by increase in temperature is observed to expedite the aging process in the molecular glasses[6-8] as well as in the soft glassy materials.[9-12] However, depending upon the precise nature of structural rearrangement present during the physical aging and the nature of low energy structures, additional effects triggered by the change in temperature may also influence this process. One of the standard ways to assess influence of other temperature dependent variables on structural recovery is to induce step-up and step-down temperature jumps and follow the evolution of the affected variable as a function of time. For examples, classical experiments by Kovacs[7] demonstrated that specific volume relaxation of polymeric glasses upon step-up and step-down change in temperature follows asymmetric paths. This behavior is explained by competition between filled volume dilation/shrinkage and free volume equilibration upon temperature change. The asymmetry arises because filled volume dilation is restricted by free volume change in the step-up jump, but filled volume shrinkage is not restricted by free volume change in the step-down jump.[13] Recently McKenna and coworkers[10] carried out similar experiments but on a soft colloidal glass composed of thermoresponsive particles of poly(N-isopropylacrylamide). These particles show rapid swelling/shrinkage dynamics and the volume of the same can be tuned by changing the temperature (diameter of particle decreases by increasing temperature). Interestingly this system also showed qualitative features of asymmetry upon step-up and down jumps in the volume fraction of particles triggered by change in temperature.

In this work we study how the structural recovery behavior of aqueous suspension of Laponite RD, a model soft glassy material, is affected by step-up and step-down temperature jumps. Primary particle of synthetic clay Laponite RD has a disk-like shape with diameter 25±2.5 nm and thickness 1 nm.[14] This material is also known to have very low size polydispersity.[15] A disk of Laponite constitutes three layers wherein two tetrahedral silica layers sandwich an octahedral magnesia layer. Isomorphic substitution of magnesium by lithium induces scarcity of positive charge such that face of Laponite particle acquires permanent negative charge.[16] In powder form Laponite particles are present in stacks with sodium ion residing in the interlayer gallery. After dispersing in water, clay layers swell and owing to osmotic pressure gradient, sodium ions dissociate in water thereby exposing negatively charged faces.[17] At pH 10, edge of Laponite particle has weak positive charge.[18] Overall three types of interactions are present among the Laponite particles in the aqueous media namely,[5] repulsion among the faces, attraction between the edge and the face, and van der Waals attraction among the particles. These interactions are responsible for the soft glassy nature of the Laponite suspension and influence its phase behavior and structural recovery.

## II. Experimental Procedure

We procured Laponite RD® from Southern Clay Products Inc. White powder of Laponite was dried at 120°C for 4 hours before mixing with ultrapure water having pH 10. A detailed procedure to prepare Laponite suspension has been described elsewhere.[5] Freshly prepared unfiltered 2.8 weight % Laponite suspension was preserved in air sealed polypropylene bottle for 3 months before using the same. The rheological experiments were performed using MCR 501 rheometer (Couette geometry: outer diameter 5.4 mm with gap of 0.2 mm). In a typical experiment sample was shear melted using oscillatory strain ($\gamma_0$ =7000 and frequency 0.1 Hz for 20 min). Subsequently oscillatory shear stress ($\sigma$=1 Pa, frequency 0.1 Hz) was applied to monitor evolution of viscoelastic behavior in the isothermal as well as in the temperature step change experiments. We verified that application oscillatory flow field with stress magnitude of 1 Pa does not affect the aging process. We also measured ionic conductivity of Laponite suspension samples using Eutech Cyberscan CON 6000 conductivity meter with 4 cell electrode (range 0-500 mS and temperature range 0-70 °C). We measured conductivity at constant temperatures from 1 to 40°C as well as in temperatures step change experiments. We carried out the conductivity experiments in the rheometer cell itself so as to identically match the temperature change profiles in the rheology experiments.

## III. Results and Discussion

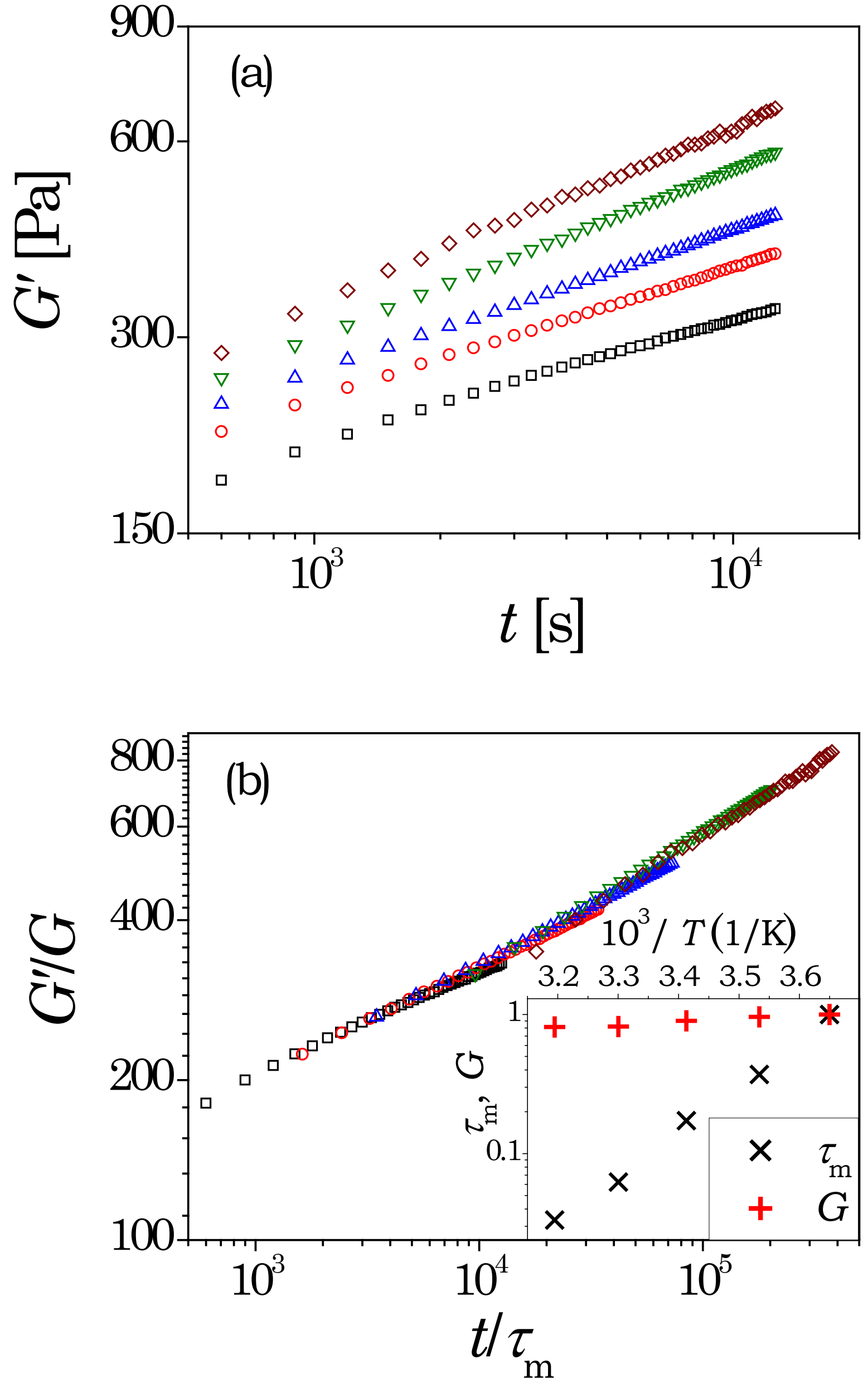

**Figure 1.** (a) Evolution of $G'$ as a function of time for different temperatures (diamonds 40°C, down triangles 30°C, up triangles 20°C, circles 10°C and squares 1°C. The figure (b) shows superposition of the evolution obtained at different temperatures. The inset shows variation of shift factors as a function of inverse of temperature required to obtain the superposition.

In figure 1 we plot evolution of elastic modulus as a function of time elapsed since stopping the shear melting (aging time) at five temperatures for Laponite suspension. It can be seen that the evolution of elastic modulus shifts to lower aging times for experiments carried out at higher temperatures. The self similar curvature of $G'$ allows formation of superposition mainly by horizontal shifting. In a glassy material, constituents of the same are kinetically arrested in physical cages formed by neighbors, which allow only a restricted access to its phase space causing ergodicity breaking. In the process of physical aging material explores its phase space and progressively attains lower values of free energy as a function of time. If we represent individual physical cages as energy wells, lowering of free energy as a function of time is equivalent to increase in well depth. If $E$ is average energy well depth, then through scaling arguments elastic modulus of a material can be written as: $G' \propto E/b^3$, where $b$ is characteristic length.[19] Furthermore, the timescale associated with the process of aging, also known as microscopic timescale ($\tau_m$), determines the rate at which average energy well depth $E$ increases: $E = E(t/\tau_m)$. Owing to faster thermal motions, $\tau_m$ is expected to decrease with increase in temperature. Increase in $G'$ can be therefore considered as an indication of decrease in free energy,[5] and time dependence of the same can be given by:

$$G'(T,t) = G(T)\, g(t/\tau_m). \qquad (1)$$

If we introduce $\tilde{G}' = G'/G(T)$ and $x = t/\tau_m$, equation (1) can be simply written as:

$$\tilde{G}' = g(x). \qquad (1a)$$

The superposition shown in figure 1b essentially represents function form of equation (1a) with $G$ and $\tau_m$ as vertical and horizontal shift factors respectively. We arbitrarily set $G(T_R)$=1 Pa and $\tau_m(T_R)$=1 s, where $T_R$ is reference temperature ($T_R$=1°C) as shown in the inset of figure (1b). $\tau_m$ is observed to increase with $1/T$ ($\tau_m$ can be seen to be following Arrhenius dependence, however the experimentally explored range of $1/T$ is very limited). The vertical shifting is necessary as increase in temperature is observed to decrease modulus in addition to its effect on $\tau_m$. Vertical shift factor $G(T)$ was always closer to unity as shown in the inset.

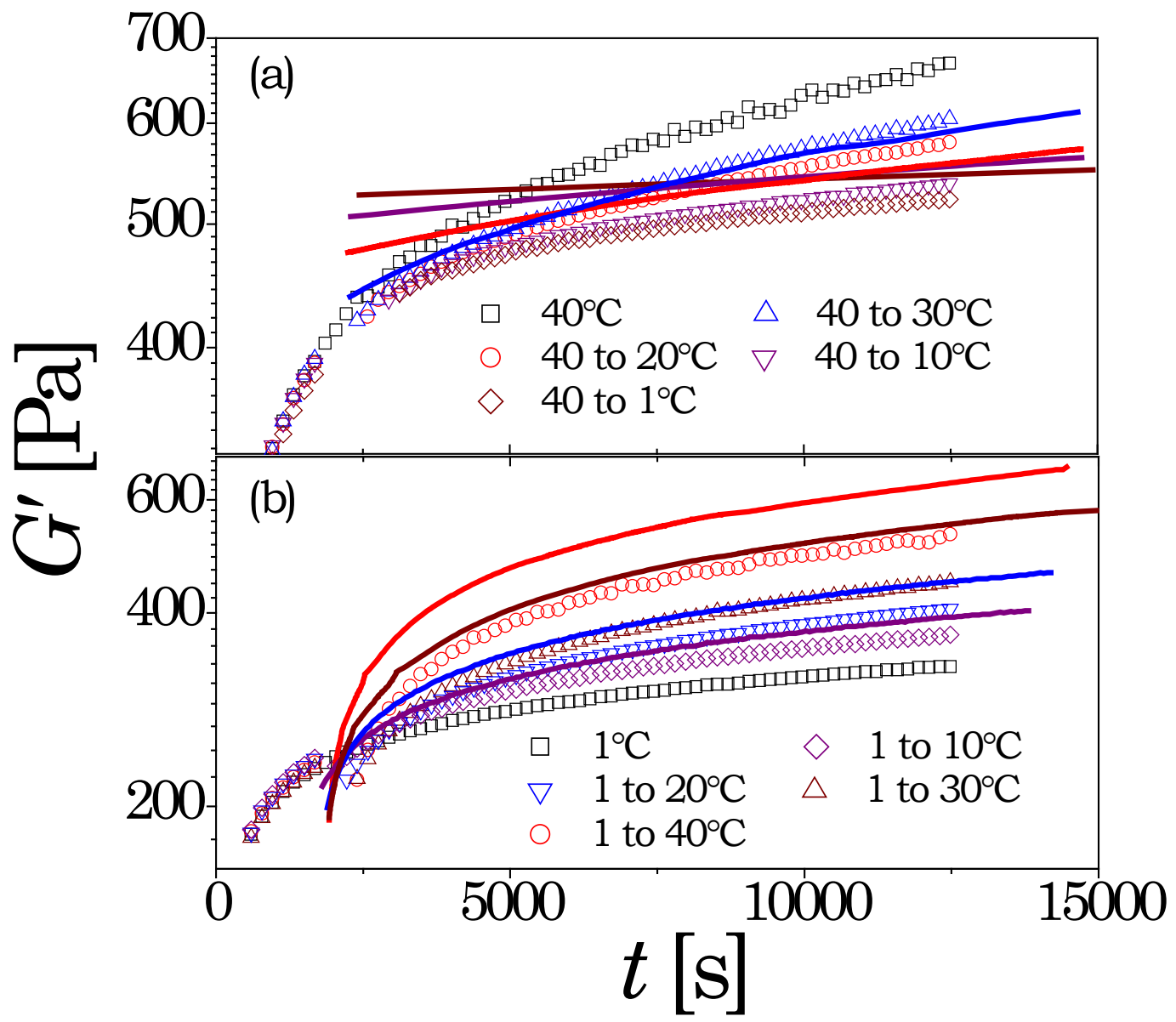

**Figure 2.** Evolution of $G'$ upon step jump in temperature at 1800 s. The top figure (a) describes the behavior for step-down jump from 40°C to mentioned lower temperatures, while the bottom figure (b) describes step-up jump from 1°C to mentioned higher temperatures. Lines passing through the data represent the predictions of equations (2) and (3) that use the isothermal aging data.

In figure 2, we plot effect of step-up and step-down temperature jump on evolution of $G'$ as a function of time. In both the cases temperature was changed at 1800 s. We have omitted the $G'$ data associated with the transient in temperature (For complete raw data refer to the Appendix). Figure 2(a) shows that greater the decrease in temperature is, lower is the rate of evolution of $G'$. Alternatively, figure 2(b) shows that larger the increase in temperature is, faster is evolution of $G'$. This scenario is described more clearly in figure 3, where evolution of $G'$ at 1, 20 and 40°C is shown in addition to that of associated with 1 to 20°C and 40 to 20°C step change. It can be seen than both the step change data approach 20°C evolution curve. In figures 2 and 3, immediately after the step increase in temperature, $G'$ can be seen to be decreasing below the isothermal aging curve of previous temperature. This is due to inverse dependence of $G'$ on temperature. However, since aging rate is greater at high temperature it soon crosses the isothermal aging curve of previous temperature. Equivalently $G'$ is expected to increase immediately after decrease in temperature; however this behavior has been masked by transient in temperature, for which we have not reported the data (Refer to Appendix for more details).

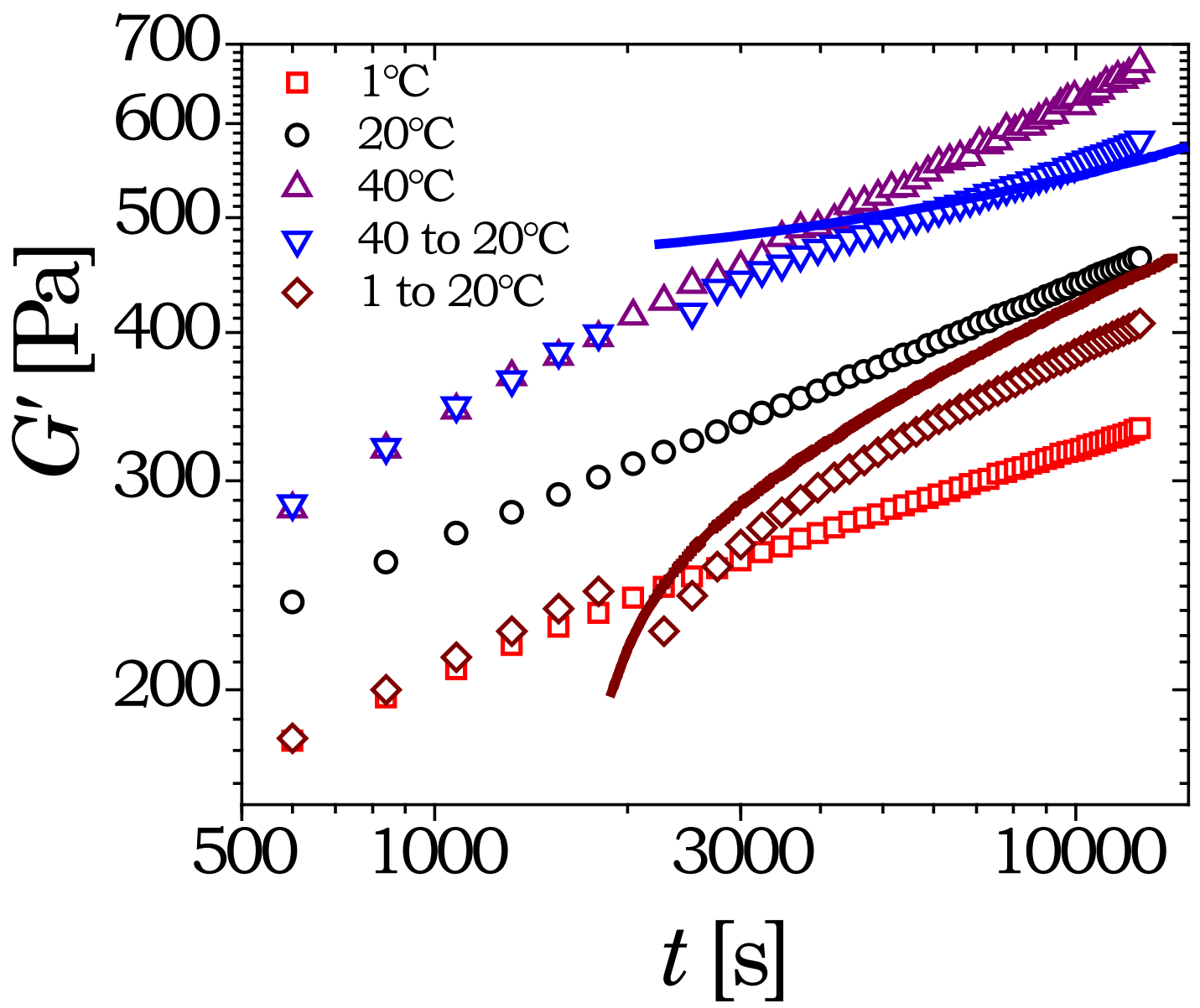

**Figure 3.** Evolution of $G'$ at 1, 20 and 40°C and after step-down jump (40 to 20 °C) and step-up jump (1 to 20°C) at 1800 s. The lines passing through the data represent predictions using isothermal evolutions and equations (2) and (3).

An assumption that $G'$ follows equation (1) suggests that change in temperature does not change the path of aging, but only the rate at which aging takes place. If such scenario exists, knowledge of evolution of $G'$ at one temperature, along with the dependences of $\tau_m$ and $G$ on $T$, can be used to predict not just evolution at any other temperature but also the evolution associated with a step change. Let us assume that step change in temperature from $T_1$ to $T_2$ was introduced at time $t_1$. However, on $\tilde{G}'$ verses $x$ ($G'/G(T)$ verses $t/\tau_m$) scale the material will continue to age on the same path, except with a different rate beyond time $t_1$. Therefore the state of the material undergoing aging at temperature $T_1$ for time $t_1$ is same as at time $t_2$ if aging would have been carried out at temperature $T_2$ since the beginning ($x = t_1/\tau_{m1}$ $= t_2/\tau_{m2}$, where $\tau_{mi}$ is associated with temperature $T_i$). Therefore, $t_2$ is given by:

$$t_2 = \left(\tau_{m2}/\tau_{m1}\right)t_1 . \qquad (2)$$

The value of $\tilde{G}'$ at time $t$ after the temperature step jump ($t > t_1$) can then simply be represented by:

$$\tilde{G}' = g(x_1 + x_2) , \qquad (3)$$

where $x_1 = t_1/\tau_{m1}$ and $x_2 = \left(t - t_1\right)/\tau_{m2}$. Evolution of $G'$ after the step jump can then be written in terms of $\tau_m$ associated with the changed temperature using equations (2) and (3) to yield:

$$G' = G'\left(T_2, t - \left(t_1 - t_2\right)\right) = G\left(T_2\right) g\left(\left[t - \left(t_1 - t_2\right)\right]/\tau_{m2}\right) \text{ ... for } t \geq t_1 . \qquad (3a)$$

Above analysis suggests that, the knowledge of evolution of $G'$ under isothermal conditions and shift factors at different temperatures should facilitate prediction of evolution after the step change through equation (3). We therefore use data presented in figure 1, to predict the evolution of $G'$ after the step change in temperature in figures 2 and 3. The prediction has been described using thick lines. We have used $t_1$=1920s for the step-up data and 2200s for the step-down data, which is different from actual value of $t_1$=1800 s owing to finite time required to complete the step change. It can be seen that equations (2) and (3) significantly over predict the evolution of $G'$ subsequent to the step-up temperature change. On the other hand predictions of step-down temperature change associated with 40°C to 20°C, 10°C and 1°C cross each other and do not follow the curvatures of the respective experimental data. Interestingly prediction of 40°C to 30°C comes closest to the experimental data.

The above discussion suggests that assumption of equation (1) results in an excellent superposition (figure 1); but its logical extension, which leads to equation (3), fails to predict the evolution subsequent to the temperature change. Consequently, this result indicates that on $G'/G(T)$ verses $t/\tau_m$ scale the material will not continue to age on the same path upon the change in temperature. This further suggests possibility of physicochemical changes to the Laponite suspension upon temperature change which may be irreversible over the experimental timescales.

Very recently it has been reported that ionic conductivity of Laponite suspension increases with increase in temperature.[5] Ionic conductivity of aqueous Laponite suspension originates from NaOH used to maintain pH 10 of water and the sodium counterions ($Na^+$) associated with Laponite particles. Ionic conductivity of ultrapure water having pH 10 (maintained by adding NaOH) is around 20 μS/cm. Therefore increase in conductivity beyond this is due to dissociation of the $Na^+$ counterions from Laponite particles. In order to study counterion ($Na^+$) dissociation behavior of Laponite suspension upon temperature change, we measured ionic conductivity of the same under isothermal conditions as well as upon temperature step change. The behavior of conductivity variation is shown in figure 4. Typically the measured values of conductivity of the suspension are over 40 times higher than the value associated with water having pH 10. Therefore major contribution to ionic conductivity is from counterions. It should be noted that before measuring the ionic conductivity, suspension samples were mechanically rejuvenated at the respective temperatures. The conductivity reported in figure 4 is under quiescent conditions and measured as a function of time after the mechanical rejuvenation was stopped. It can be seen from figure 4 that conductivity of suspension kept under isothermal conditions remains constant over the duration of experiment, but indeed increases with increase in temperature. We also plot conductivity upon step change in temperature in the same plots without disturbing the material as is the case with rheology experiments. It can be seen that upon temperature step-up and step-down change, conductivity respectively increases and decreases subsequent to the change, but does not reach the same value associated with the isothermal measurements of the changed temperature. In addition, for smaller changes in temperature the difference of isothermal conductivity value and that of upon temperature change is small, but increases with increase in the magnitude of temperature change.

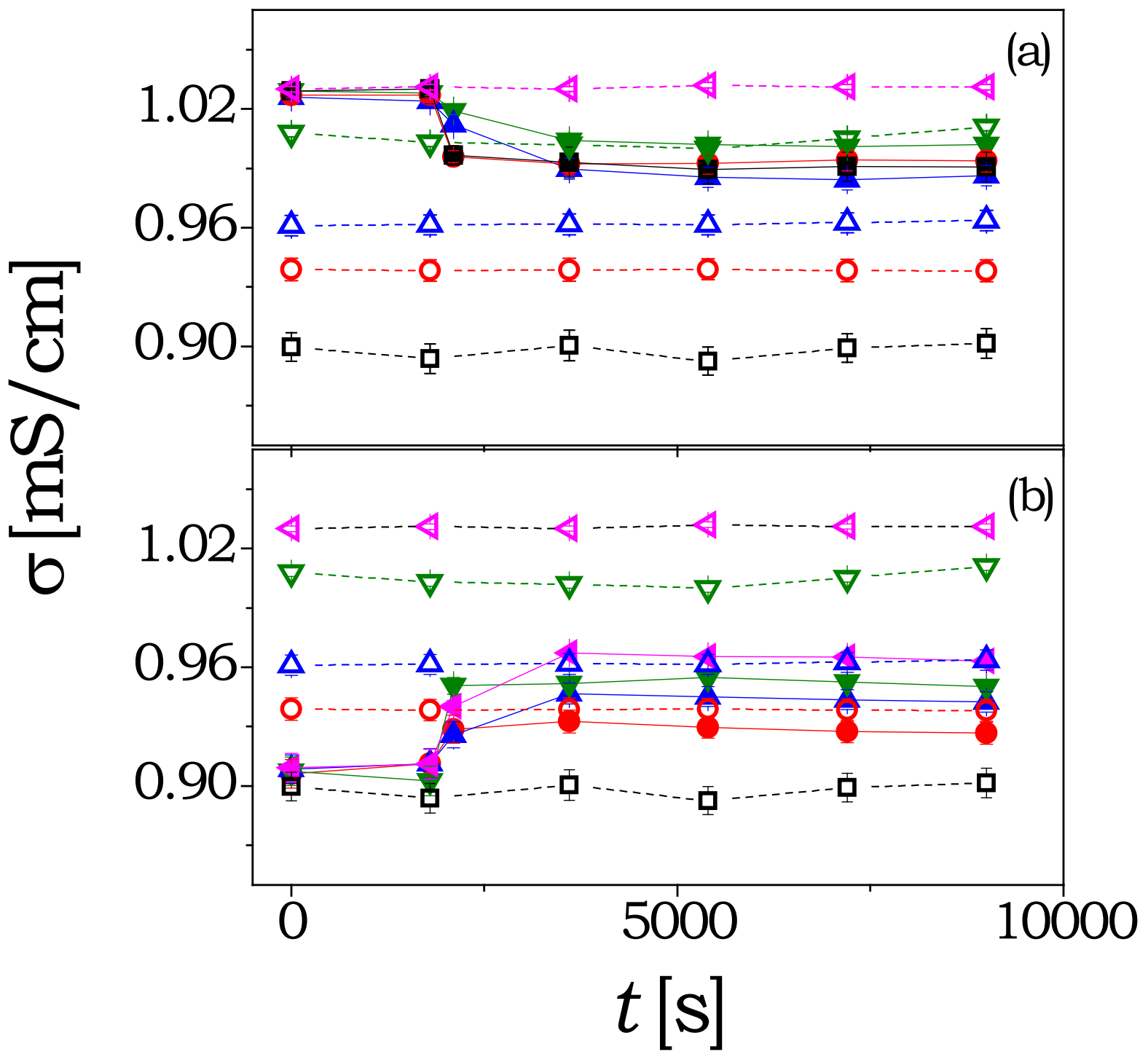

**Figure 4.** Ionic conductivity of Laponite suspension is plotted as a function of time under isothermal conditions (open symbols) and following the temperature change conditions (filled symbols). The temperature change was carried out at 1800 s. In both the figures open symbols from bottom to top: 1°C, 10 °C, 20°C, 30°C, and 40°C. Filled

symbols in figure (a) represent temperature down jump: black squares 40-1°C, red circles 40-10°C, blue up triangles 40-20°C, green down triangles 40-30°C. Filled symbols in figure (b) represent temperature up jump: red circles 1-10°C, blue up triangles 1-20°C, green down triangles 1-30°C, and magenta left triangles 1-40°C.

Usually subsequent to the temperature change ionic conductivity (or the concentration of counterions) is expected to equilibrate with respect to the changed temperature. In case of increase in temperature more dissociation of counterions and their diffusion away from the particle surface is expected. On the other hand with decrease in temperature part of the counterions are expected to recombine with the faces of Laponite particle. Therefore, the plausible reason behind the difference in counterion concentration under isothermal conditions and that of after temperature change may be the soft solid like consistency associated with Laponite suspension at the time of temperature change. In the conductivity experiments we ensure thermal equilibrium before the mechanical rejuvenation is performed. During the mechanical rejuvenation the material is in liquid state and has low viscosity, consequently the mobility of the counterions is high and equilibrium value of ionic conductivity associated with the corresponding temperature is achieved. However, at the time of temperature change the material is not in the liquid state but in an apparent solid state ($G'$ is much above 200 Pa and significantly greater than $G''$) and owing to which, reduction in mobility of the counterions may take place. Such reduction in the mobility of counterions leads to restricted diffusion of counterions hindering achievement of equilibration of conductivity in the temperature step change experiments. Consequently, in temperature step-down experiments conductivity remains at high value even at reduced temperature, while in step-up experiments conductivity remains at low value though the temperature is high. We believe that this phenomenon is responsible for asymmetry in the rheological behavior in the up and down temperature jump experiments.

Concentration of counterions is known to profoundly influence interactions among the Laponite particles as it directly affects the Debye screening length ($1/\kappa$) as well as the surface potential ($\Phi_0$) associated with the faces of Laponite particles. Typically dependence of Debye screening length on concentration of $Na^+$ ions is given by: $1/\kappa \sim n_{Na}^{-0.5}$,[20] while that of surface potential is given by: $\Phi_0 \sim \left(n_{Na} - n_0\right)$,[5] where $n_0$ is number density of $Na^+$ ions due to the sources other than Laponite (such as NaOH). Shahin and Joshi[5] recently analyzed interactions among the Laponite particles using DLVO theory and observed that enhancement in conductivity owing to counterions increases the height of repulsive energy barrier, however decreases the width of the same when particles approach each other in a face-to-face fashion. Therefore, due to difference in conductivity associated with isothermal and temperature change experiments, equation (2) and in turn equation (3) cannot be applied to the same.

From the comparison of the predictions of equation (3) and the experimental data shown in figure 2, it is plausible to conclude that decrease in concentration of $Na^+$ ions causes increase in $G'$. This is more apparent for the comparison of step-down temperature change experiments (for step-up temperature change experiments this effect is observable over a small duration of time). In figure 2, prediction subsequent to step-down temperature jump shows higher increase in $G'$ for greater decrease in $T$ as model considers isothermal data for which concentration of $Na^+$ ions substantially decreases with decrease in $T$. As time passes, owing to higher rate of aging at higher $T$, the predictions cross each other. Qualitatively similar behavior is also observed for step-up temperature jump predictions with cross over occurring soon after the temperature change. In reality, however, decrease in temperature does not decrease concentration of $Na^+$ ions to the level of isothermal values. Consequently, increase in $G'$ in step-down experiments and decrease in $G'$ in step-up experiments is significantly smaller than that predicted by the model.

Interestingly the conductivity after the temperature step–down jump from 40°C to 30°C is very similar to that of isothermal conductivity associated with 30°C as shown in figure 4 (a). As a result, we can apply equations (2) and (3) to the $G'$ evolution data. Consequently as shown in figures 2, the prediction of the behavior upon temperature change for 40°C to 30°C comes closest to that of experimental data.

In soft glassy materials the interparticle interactions are primarily responsible for their inability to achieve equilibrium.[1] Therefore for those soft glassy materials, where the constituents of the same bear dissociable counterions and charged surfaces, hindered mobility of the ions cannot be ruled out. For such systems aging behavior upon step temperature change can be similar to that of observed in the present work. The present results therefore suggests apparent similarity with the observation of asymmetric structural recovery seen in the in the polymeric glasses as reported by Kovacs[7] and for thermo-sensitive soft glassy paste as observed by McKenna and coworkers.[10] In all these cases asymmetry can be attributed to other nonlinear physical effects induced due to change in temperature. However these other nonlinear effects originate from kinetically arrested nature of the material. In the first case it is asymmetric dissociation of counterions upon change in temperature while in the latter cases it is asymmetry associated with filled volume expansion/contraction triggered by the temperature change.

## IV. Conclusion

In this work, we monitor evolution of $G'$ of a shear rejuvenated soft glassy Laponite suspension under isothermal as well as step temperature change experiments. It is typically observed that structural recovery described by increase in $G'$ is faster at greater temperatures, and speeds up or slows down respectively for the step-up and step-down changes in temperature. Superposition of isothermal evolution curves of $G'$ at different temperatures supports an assumption that only the rate of structural recovery and not the path depends on temperature. However, when temperature change was carried out in the solid state, this assumption fails to predict the evolution of $G'$ following step change from the isothermal data suggesting temperature change induces such effects that are irreversible over the observation timescales. We also measure the ionic conductivity of the Laponite suspension which indicates concentration of dissociated counterions. For experiments carried out at constant temperatures, conductivity is observed to be higher at greater temperatures. However during the temperature jump experiments, owing to restricted mobility of the ions at the time of jump when the material is in solid state, change in conductivity does not reach the isothermal conductivity values associated with the changed temperature. We feel that this nonlinearity associated with conductivity change causes rheological response to be asymmetric.

**Acknowledgement:** We would like to thank Department of Atomic Energy – Science Research Council (DAE-SRC) for financial support. We also thank an anonymous reviewer for thoughtful comments that led to substantially improved analysis of the experimental data.

# Appendix

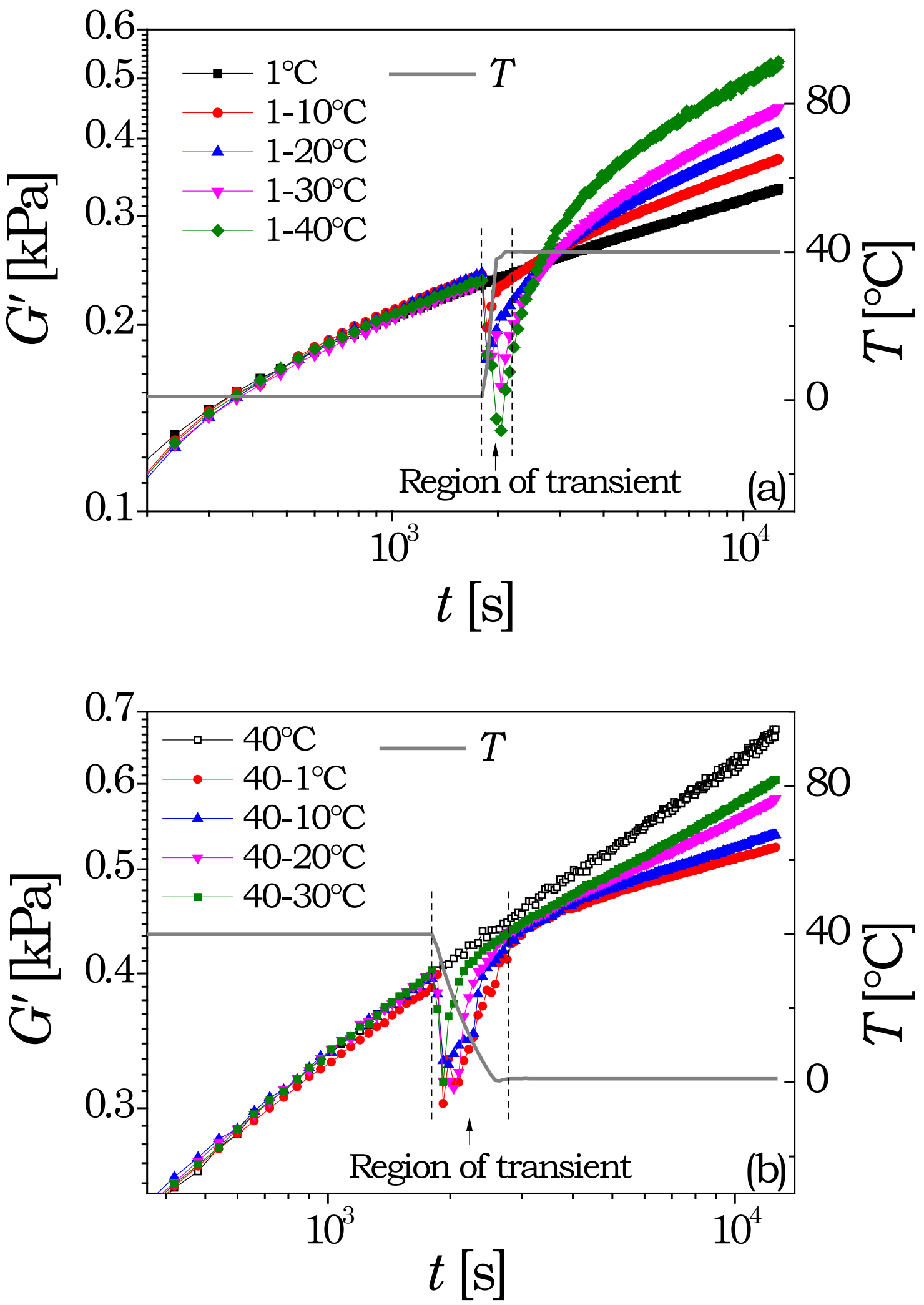

**Figure S1.** Raw data of $G'$ evolution upon temperature step up (a), and step down (b) jump for Laponite suspension. The corresponding change in temperature for one case each is also plotted in the respective figures. It can be seen that time required to overcome transient in temperature decrease (≈1000 s) is higher than that required for temperature increase (≈300 s). During such transients, since properties of the material change rapidly over a duration of a single cycle to monitor $G'$, response ceases to be harmonic. This induces errors in the estimation of $G'$. Therefore we have completely omitted the data associated with the transient in the manuscript.